
\hfuzz 4pt

\setbox0=\hbox{\raise2pt\hbox{$\chi$}}

\def\Tr{{\rm Tr}}
\def\tr{{\rm tr}}
\font\titlefont=cmbx10 scaled\magstep1

\magnification=\magstep1

\null
\line{\hfill S.I.S.S.A. 98/95/EP}
\line{\hfill \tt hep-th 9508157}
\vskip 2cm
\centerline{\titlefont QUANTUM MECHANICAL BREAKING}
\smallskip
\centerline{\titlefont OF LOCAL GL(4) INVARIANCE}
\vskip 2cm
\centerline{\bf R. Floreanini}
\smallskip
\centerline{Istituto Nazionale di Fisica Nucleare, Sezione di Trieste}
\centerline {Dipartimento di Fisica Teorica,
Universit\`a di Trieste}
\centerline{Strada Costiera 11, 34014 Trieste, Italy}
\vskip 0.5cm
\centerline{and}
\vskip 0.5cm
\centerline{\bf R. Percacci}
\smallskip
\centerline{International School for Advanced Studies, Trieste, Italy}
\smallskip
\centerline{and}
\smallskip
\centerline{Istituto Nazionale di Fisica Nucleare, Sezione di Trieste}
\vskip 2.5cm
\centerline{\bf Abstract}
\smallskip\midinsert\narrower\narrower\noindent
We consider the gravitational coupling of a scalar field,
in a reformulation of General Relativity exhibiting local
$GL(4)$ invariance at the classical level.
We compute the one--loop contribution of the scalar to the
quantum effective potential of the vierbein and find that it
does not have $GL(4)$ invariance.
The minima of the effective potential occur for a vierbein which is
proportional to the unit matrix.
\endinsert
\vfil\eject

\leftline{\bf 1. Introduction}
\smallskip
\noindent
In recent papers [1-3] we have discussed the quantum mechanical
breaking of scale invariance in quantum gravity.
One begins from a classical action which depends on the
metric $g_{\mu\nu}$ only through the combination
$$
\tilde g_{\mu\nu}=\rho^2 g_{\mu\nu}\ ,\eqno(1.1)
$$
where $\rho$ is a scalar field in
the gravitational sector of the theory, called the dilaton.
The classical action is then invariant under the local scale
transformations
$$
g_{\mu\nu}\to g'_{\mu\nu}=\Omega^2g_{\mu\nu}\ \ ,\qquad\qquad
\rho\to\rho'=\Omega^{-1}\rho\ .
\eqno(1.2)
$$
If in the functional integral the measure is defined with the
metric $g$, rather than $\tilde g$,
the functional integral,
and hence in particular the effective potential, will be a
function of $g$ and $\rho$ separately and therefore invariance
under (1.2) will be broken.

Using traditional field theoretic methods we have computed
the renormalized effective potential for $\rho$ and found it
to be of the Coleman--Weinberg type [1, 2].
Using the so-called average effective action we have also
computed the renormalization group flow of various quantities
of interest, in particular of the v.e.v. of the dilaton [2, 3].
This is relevant to the quantization of gravity, since the v.e.v.
of the dilaton is proportional to Newton's constant.

Here we report on a generalization of these results, where
local scale invariance is enlarged to local $GL(4)$ invariance.
Instead of the dilaton field $\rho$ one has a matrix--valued
field $\theta^\mu{}_\nu$, and the classical action depends
only on the combination
$$
\tilde g_{\mu\nu}=\ \theta^\rho{}_\mu\, \theta^\sigma{}_\nu\,
g_{\rho\sigma}\ .\eqno(1.3)
$$
The theory is invariant under the local $GL(4)$ transformations
$$
\eqalignno
{g_{\mu\nu}(x)&\mapsto{g^\prime}_{\mu\nu}(x)=
\Lambda^\rho{}_\mu(x)\, \Lambda^\sigma{}_\nu(x)\,
g_{\rho\sigma}(x)\ ,&(1.4a)\cr
\theta^\rho{}_\mu(x)&\mapsto{\theta^\prime}^\rho{}_\mu(x)=
\Lambda^{-1 \rho}{}_\sigma(x)\, \theta^\sigma{}_\nu(x)\ ,&(1.4b)\cr}
$$
which leave $\tilde g_{\mu\nu}$ invariant.
The dilatonic theory is recovered if we assume that
$\theta^\mu{}_\nu=\rho\,\delta^\mu_\nu$.
The local $GL(4)$ transformations specialize to the scale
transformations (1.2) when
$\Lambda^\rho{}_\mu=\Omega\,\delta^\rho_\nu$.

The reformulation of General Relativity or any other theory
of gravity in this $GL(4)$--invariant way has been discussed
in [4] and independently in [5], where
the connection with Weyl's geometry was emphasized.

In addition to local $GL(4)$ invariance one also considers
diffeomorphisms. There are two ways of realizing the diffeomorphism
group on the fields [1]. The first is given by
$$
\eqalignno
{g_{\mu\nu}(x)&\mapsto{g^\prime}_{\mu\nu}(x^\prime)=
{\partial x^\rho\over\partial {x'}^\mu}
{\partial x^\sigma\over\partial {x'}^\nu}
g_{\rho\sigma}(x)\ ,&(1.5a)\cr
\theta^\rho{}_\mu(x)&\mapsto{\theta^\prime}^\rho{}_\mu(x^\prime)=
\,{\partial {x'}^\rho\over\partial x^\sigma}
\theta^\sigma{}_\nu(x)
{\partial x^\nu\over\partial x^{\prime \mu}}\ .&(1.5b)\cr}
$$
In this realization all fields are transformed as tensors
on all indices.

The second realization consists of the first, followed by
a local $GL(4)$ transformation with parameter
$\Lambda^\mu{}_\nu={\partial {x'}^\mu\over\partial x^\nu}$:
$$
\eqalignno
{g_{\mu\nu}(x)&\mapsto{g^\prime}_{\mu\nu}(x^\prime)=
g_{\mu\nu}(x)\ ,&(1.6a)\cr
\theta^\rho{}_\mu(x)&\mapsto{\theta^\prime}^\rho{}_\mu(x^\prime)=
\,\theta^\rho{}_\nu(x)
{\partial x^\nu\over\partial x^{\prime \mu}}\ .&(1.6b)\cr}
$$
In both cases the metric field $\tilde g_{\mu\nu}$ transforms as usual.

There are two possible geometrical interpretations of this theory.
In [4] we regarded the first index of $\theta$ and the indices on $g$
as internal indices. In this interpretation, (1.6) describes the
effect of a diffeomorphism on the fields.
In the present paper we will adopt another interpretation, namely
we treat all indices as coordinate indices in the tangent bundle.
In this interpretation the action of a diffeomorphism is described by
(1.5).

Just as in the case of scale invariance, local $GL(4)$
invariance will be broken in the quantum theory if the functional
measure is constructed with the metric $g$ rather than $\tilde g$ [1].
In particular, the effective potential will be a function of
$g$ and $\theta$ separately, rather than just of $\tilde g$.
It can be regarded as an effective potential for the field
$\theta$.

The question is now: is the resulting effective potential
a sensible one? If so, where are its minima?
We know that in the case $\theta^\mu{}_\nu=\rho\,\delta^\mu_\nu$
the potential for $\rho$ is bounded from below and has a
minimum for some nonzero value of $\rho$.
This is helpful information, but it does not guarantee that the
potential for $\theta$ will have the same minimum.
In this paper we will compute the contribution to the
effective potential for $\theta$ coming from the quantum fluctuations
of a scalar field.
We show that the minimum occurs when $\theta$ is a
multiple of the identity, and therefore coincides with the
minimum that was found in the dilatonic theory.

The calculation of this potential in a full theory of gravity,
taking into account the quantum fluctuations of
fermions, photons and gravitons,
is technically a much more complicated problem, but we expect
that the final results will not be qualitatively different
from the ones that we find here.

Throughout this paper we will work in the Euclidean theory.
This is just to simplify the notation:
there is no obstacle in doing the same calculations
in the Minkowskian theory.

\bigskip
\leftline{\bf 2. The contribution of a scalar field}
\smallskip
\noindent
In this paper we consider a scalar field $\phi$ coupled to gravity.
Our basic assumption is that in the classical action
(which in a quantum context describes the physics at the cutoff scale)
the scalar is minimally
coupled to the metric $\tilde g$ given in (1.1):
$$
\eqalign{
S(\phi,g,\theta)=&{1\over2}\int d^4x\sqrt{\det \tilde g}
\left[\tilde g^{\mu\nu}\partial_\mu\phi\partial_\nu\phi
+c\, \phi^2\right]\cr
=&{1\over2}\int d^4x\sqrt{\det g}\,{\det\theta}
\left[g^{\mu\nu}\theta^{-1}_\mu{}^\rho\theta^{-1}_\nu{}^\sigma
\partial_\rho\phi\partial_\sigma\phi
+c\, \phi^2\right]\ .\cr}
\eqno(2.1)
$$
Note that the scalar is coupled in a complicated nonminimal way to the
metric $g_{\mu\nu}$ and the field $\theta^\mu{}_\nu$.
We assume that $g_{\mu\nu}$ is dimensionless,
$\theta^\mu{}_\nu$ has dimension of mass, $\phi$ is
dimensionless and $c$ is dimensionless.
The physical picture underlying these choices has been discussed
extensively, in the context of the dilatonic theory, in [2, 3].

The action (2.1) is invariant under the local $GL(4)$ transformations
(1.4) and under the diffeomorphisms (1.5).
Let us assume that the background
metric $g_{\mu\nu}$ is flat. We can then choose the gauge so that
$g_{\mu\nu}=\delta_{\mu\nu}$. This breaks local $GL(4)$
to local $O(4)$ invariance and diffeomorphisms to global $O(4)$
transformations. From (1.3) we see that in this case the field
$\theta^\mu{}_\nu$ can be interpreted formally as a vierbein.
We are interested in
the effective potential for $\theta$, so we shall assume that the
matrix $\theta$ is constant.
Under these circumstances the residue of the initial local $GL(4)$
invariance consists of global $O(4)$ transformations acting only on the
first index of $\theta$, while the residue of the diffeomorphism
invariance consists of global $O(4)$ transformations
acting on the matrix $\theta^\mu{}_\nu$ by similarity
transformations. These two invariances are equivalent to
left and right $O(4)$ invariance, the right $O(4)$ being the
residue of the transformations (1.6).

Since the metric is flat we go to momentum space.
The action (2.1) then becomes:
$$
S(\phi,\theta)=
{1\over2}\det\theta\int {d^4q\over (2\pi)^4}\,
\phi(q)\,{\cal O}(q)\,\phi(-q)\ .\eqno(2.2)
$$
The operator ${\cal O}$ is given by
$$
{\cal O}(q)={\rm tr}\left(q^T\theta^{-1T}\theta^{-1}q\right)+c\ ,
\eqno(2.3)
$$
where $q$ is thought of as a column vector.

There is an issue of how to treat the factor $\det\theta$
appearing in front of the integral, which is related to the
choice of functional measure in the path integral.
Let us assume first that the measure is given (formally) by
$d\mu(\phi)=\prod_k d\phi(k)$.
The one--loop effective action for $\theta$ induced by quantum
fluctuations of the scalar field is the determinant of the
quadratic operator appearing in (2.2):
$$
\Gamma_{\rm eff}(\theta)=
\int d^4x V_{\rm eff}(\theta)=
{1\over2}\ln{\rm Det}(\det\theta\cdot{\cal O})=
{1\over2}\Tr\ln(\det\theta\cdot{\bf 1})+{1\over2}\Tr\ln{\cal O}\ ,
\eqno(2.4)
$$
where $V_{\rm eff}$ is the effective potential,
Det and Tr denote functional determinant and trace.

We have separated the contribution of the operator
$\det\theta\cdot{\bf 1}$,
which is proportional to the unit matrix in the function space.
This term is equal to
$$
{{\cal V}\over2}\ln(\det\theta)\, \delta(0)\ ,\eqno(2.5)
$$
where ${\cal V}=\int d^4x$ is the spacetime volume.
If we evaluate the functional trace in Fourier space and introduce
an UV cutoff $\Lambda$ we have
$\delta(0)=\int {d^4 q\over (2\pi)^4}={1\over32\pi^2}\Lambda^4$.

On the other hand suppose we define the measure as
$d\mu(\phi)=\prod_k d\phi(k)\,(\det\theta)^{K}$.
This is equivalent to saying that the quantum field is not
$\phi$ but rather $\varphi$, where
$$
\varphi=\phi\,(\det\theta)^K\ .\eqno(2.6)
$$
The effective action changes by the addition of a term
$$
-K\, {\cal V} \,\ln(\det\theta)\,\delta(0)\ .\eqno(2.7)
$$
In this paper we will assume that $K=1/2$,
so as to eliminate the first term on the r.h.s. of (2.4).
This choice gives a dimension of squared mass to the quantum field
$\varphi$. Instead, the canonical dimension of mass for $\varphi$ is
obtained when $K=1/4$. This is the natural choice when
$\theta^\mu{}_\nu=\rho^2\delta^\mu_\nu$ [1-3], since it simply gives:
$\Gamma_{\rm eff}={1\over2}\ln{\rm Det}(q^2+c\,\rho^2)$.
However, when the eigenvalues of $\theta^\mu{}_\nu$
are all different, the choice $K=1/4$ would in general produce, after
renormalization, additional terms in $\Gamma_{\rm eff}$ proportional
to $\ln\det\theta$; these can be eliminated with a further
finite, but non-polynomial, renormalization.

The effective potential is then given by the formula
$$
V_{\rm eff}(\theta)=
{1\over2}\int {d^4q\over (2\pi)^4}\ln{\cal O}\ .\eqno(2.8)
$$
This integral is divergent, so we regulate it with an
UV cutoff $\Lambda$. Since the integration measure
is defined with the background metric $g_{\mu\nu}=\delta_{\mu\nu}$,
the cutoff is defined by $\sum_\mu q_\mu^2<\Lambda^2$.
As we have emphasized in [1, 2] there is in principle also
the possibility of using the metric $\tilde g$ to define the
cutoff, leading to very different results.
It is the physical interpretation that dictates the
choice we are making here.

In order to evaluate (2.8) we proceed as follows. We observe
that using the residual global left and right
$O(4)$ invariances, the matrix
$\theta$ can be brought to diagonal form.
Let $M_L$ and $M_R$ be such that
$$
(M_L\cdot\theta\cdot M_R)_{\mu\nu}=\theta_\mu\delta_{\mu\nu}\ .
\eqno(2.9)
$$
(Note that $\theta_\mu$ are not the eigenvalues of the matrix
$\theta^\mu{}_\nu$.)
We then perform a change of variables $q'=M_L q$ in the integral (2.8).
The cutoff condition is clearly unchanged.
Inserting $M_R M_R^T={\bf 1}$ in the operator ${\cal O}$,
the matrix $\theta$ is diagonalized and the effective potential
becomes
$$
V_{\rm eff}(\theta)=
{1\over2}\int {d^4q\over (2\pi)^4}\ln
\left[\sum\limits_{\mu=1}^4 {q_\mu^2\over\theta_\mu^2}+c\right]\ .
\eqno(2.10)
$$
The integrand is now a function whose level surfaces are three
dimensional ellipsoids. We have to evaluate the integral of this
function over a spherical ball of radius $\Lambda$.

Note that $V_{\rm eff}$ depends on the eigenvalues
$\theta_\mu$ only through their squares and therefore
must be an even function of these variables.

\bigskip
\leftline{\bf 3. Two dimensions}
\smallskip
\noindent
In order to gain some insight in a simpler setting we shall
evaluate first the effective potential for $\theta$ induced by a
scalar field in two dimensions.
It is given by the expression
$$
V_{\rm eff}(\theta)=
{1\over2}\int {d^2q\over (2\pi)^2}
\ln\left[{q_1^2\over\theta_1^2}+
{q_2^2\over\theta_2^2}+c\right]
={1\over8\pi^2}\int\limits_0^{2\pi} d\varphi
\int\limits_0^{\Lambda} dr\, r \ln(\omega r^2+c)\ ,\eqno(3.1)
$$
where we have used polar coordinates, defined by
$q_1=r\cos\varphi$, $q_2=r\sin\varphi$,
and
$\omega(\varphi)=
{\cos^2\varphi\over\theta_1^2}+{\sin^2\varphi\over\theta_2^2}$.
One can perform first the radial integral, leading to
$$
V_{\rm eff}
={1\over16\pi^2}\int\limits_0^{2\pi} d\varphi
\left[(\Lambda^2+{c\over\omega})\ln(\Lambda^2\omega+c)
-\Lambda^2-{c\ln c\over\omega}\right]\ ,\eqno(3.2)
$$
For $\Lambda^2\omega\gg 1$ one can expand the logarithm and
discard terms of order ${1\over\Lambda^2}$.
Integrating over the angle we then get:
$$
V_{\rm eff}=
{1\over4\pi}\left[\Lambda^2
\left(\ln{\Lambda(|\theta_1|+|\theta_2|)\over 2|\theta_1\theta_2|}
-{1\over 2}\right)
+c|\theta_1\theta_2|
\left(\ln{2\Lambda\over\sqrt c\, (|\theta_1|+|\theta_2|)}
+{1\over 2}\right)\right]\ .
\eqno(3.3)
$$
Note that this is indeed an even function of $\theta_\mu$,
as expected.
If we introduce a mass parameter $\mu\ll\Lambda$,
playing the role of renormalization
point, we can rewrite (3.3) as
$$
V_{\rm eff}=
{1\over4\pi}\left[\Lambda^2
\ln{\mu(|\theta_1|+|\theta_2|)\over 2|\theta_1\theta_2|}
+c|\theta_1\theta_2|\ln{\Lambda\over\mu}
+c|\theta_1\theta_2|
\left(\ln{2\mu\over\sqrt c\, (|\theta_1|+|\theta_2|)}
+{1\over 2}\right)\right]\ ,
\eqno(3.4)
$$
where some $\theta$--independent terms have been dropped.
This way of writing reveals the presence of quadratic and
logarithmic divergences.
The unusual feature of this result is that the divergent terms depend
in a nonpolynomial way on $\theta_1$ and $\theta_2$.
This feature of quantum gravity had been noted long ago [6].

To cancel these divergences one needs therefore nonpolynomial
counterterms.
Let us choose a renormalization scheme in which the divergences are
exactly cancelled. The effective potential is then given by the finite term
in (3.4). The renormalized, finite, effective action can be written
$$
S_{\rm fin}(\theta)=
\int d^2x V_{\rm fin}(\theta)=
-{c\over8\pi}\int d^2x|\det\theta|
\left\{\ln{c\over 4\mu^2}
\left[\tr\theta^T\theta+2\det\theta(1-{\rm
sign}\det\theta)\right]-1\right\}
\ .\eqno(3.5)
$$
Note that the effective potential is invariant
under the left and right $O(2)$ transformations.

The effective potential (3.5) has an absolute maximum
for $\theta_1=\theta_2=\mu/\sqrt c$.
The fact that it is unbounded from below is a
standard problem with two dimensional theories and will
not concern us here.
The purpose of this calculation was simply to acquire
some experience with the evaluation of integrals of the
type (2.10).

\bigskip
\leftline{\bf 4. Four dimensions}
\smallskip
\noindent

Introducing polar coordinates
$$
\eqalign{
q_1\equiv&\,r\hat q_1=\,r\sin\chi\sin\alpha\cos\varphi\cr
q_2\equiv&\,r\hat q_2\,=r\sin\chi\sin\alpha\sin\varphi\cr
q_3\equiv&\,r\hat q_3\,=r\sin\chi\cos\alpha\cr
q_4\equiv&\,r\hat q_4\,=r\cos\chi}\eqno(4.1)
$$
the integral (2.10) becomes
$$
V_{\rm eff}(\theta)=
{1\over 32\pi^4}\int d\Omega
\int\limits_0^\Lambda dr r^3
\ln\left[r^2\omega+c\right]\ ,\eqno(4.2)
$$
where now
$$
\omega(\chi,\alpha,\varphi)
=\sum\limits_{\mu=1}^4 {\hat q_\mu^2\over\theta_\mu^2}=
{\cos^2\chi\over \theta_4^2}
+\sin^2\chi\left({\cos^2\alpha\over\theta_3^2}
+\sin^2\alpha
\left({\cos^2\varphi\over \theta_1^2}
+{\sin^2\varphi\over \theta_2^2}\right)\right) \eqno(4.3)
$$
and
$$
\int d\Omega=\int\limits_0^{2\pi}d\varphi
\int\limits_0^\pi d\alpha\sin\alpha
\int\limits_0^\pi d\chi\sin^2\chi\ .
$$
One can perform the radial integration, leading to
$$
V_{\rm eff}(\theta)=
{1\over128\pi^4}\int d\Omega
\left[\left(\Lambda^4-{c^2\over\omega^2}\right)\ln(\Lambda^2\omega+c)
-{\Lambda^4\over2}+{\Lambda^2 c\over \omega}
+{c^2\over\omega^2}\ln c\right]\ .\eqno(4.4)
$$
Unlike the two-dimensional case, the angular integration is too
complicated to be performed exactly. It can be done explicitly
in the case when two eigenvalues are equal (for example
$\theta_1$ and $\theta_2$), since then one angular integration becomes
trivial. This calculation is not very illuminating and will not
be reported here.

When the cutoff $\Lambda$ becomes very large,
expanding the logarithms as in the previous section we
arrive at the expression
$$
\eqalign{
V_{\rm eff}(\theta)=
{1\over128\pi^4}\int d\Omega &
\Biggl[\Lambda^4\left(\ln{\Lambda^2\over\mu^2}-{1\over2}\right)
+\Lambda^4\ln(\mu^2\omega)\cr
& +2c{\Lambda^2\over \omega}
+{c^2\over\omega^2}\ln{\Lambda^2\over\mu^2}
+{c^2\over\omega^2}\left(\ln{c\over\mu^2\omega}-{1\over2}\right)
\Biggr]\ .\cr}\eqno(4.5)
$$
The first term is field--independent and will not be considered
further; the next three terms are quartically, quadratically
or logarithmically divergent.
These divergent terms are all nonpolynomial in $\theta_\mu$.
Finally, the last term is finite.

We will adopt a renormalization prescription which amounts
simply to discarding all the divergent terms.
The renormalized effective potential is therefore
$$
V_{\rm fin}(\theta)=
{1\over128\pi^4}\int d\Omega
\left[{c^2\over\omega^2}\left(\ln{c\over\mu^2\omega}-{1\over2}\right)
\right]\ .\eqno(4.6)
$$

Even though we cannot write a closed expression for the potential,
we can study it by first taking derivatives with respect to
$\theta_\mu$'s, setting the $\theta_\mu$'s to some desired value
and then evaluating the integrals.
In this way we can compute the Taylor expansion of $V_{\rm eff}$
around some point.

The point we choose is $\theta_1=\theta_2=\theta_3=\theta_4=\mu/\sqrt c$,
which is known to be the minimum of the potential when all
the $\theta_\mu$'s are equal.
(Because of the even parity of the potential, one could also
individually change the signs of the $\theta_\mu$'s
without affecting the following arguments).

Taking the first derivative of (4.6) with respect to $\theta_\mu^2$,
we find after a little algebra
$$
{\partial V_{\rm fin}\over \partial\theta_\mu^2}=
{1\over64\pi^2}{c^2\over\theta_\mu^4}
\int d\Omega\,
{\hat q_\mu^2\over\omega^3}\ln{c\over\omega}\ .\eqno(4.7)
$$
Evaluating at the specified point gives zero.
Therefore the chosen point is a stationary point of the
effective potential.
To see what kind of stationary point, we evaluate the second
derivatives at the point
$$
{\partial^2 V_{\rm fin}\over \partial\theta_\mu^2\partial\theta_\nu^2}
\Bigg |_{\theta_\nu=\mu/\sqrt c}=
{c^2\over64\pi^4}
\int d\Omega
\,\hat q_\mu^2\hat q_\nu^2=
{c^2\over 768\pi^2}
\left(2\delta_{\mu\nu}+1\right)\ .
\eqno(4.8)
$$
The matrix in parentheses has eigenvalues 6, 2, 2 and 2 and therefore
the stationary point is a local minimum. One can check
that it is the absolute minimum of the effective potential, which
therefore coincides with the minimum found in the dilatonic theory [1, 2],
where the field $\theta$ was taken to be proportional to the unit
matrix from the beginning.

The effective potential can be approximated in the neighbourhood
of the minimum by
$$
V_{\rm fin}={1\over 64\pi^2}
\left[-2\mu^4-{c\mu^2\over 2}\tr\theta^T\theta+
{c^2\over24}\left((\tr\theta^T\theta)^2+2\tr(\theta^T\theta)^2\right)\right]\ ,
\eqno(4.9)
$$
where we have used a matrix notation for $\theta^\mu{}_\nu$.
This is exactly the type of potential that was studied in [7, 8].
There, this kind of potential was constructed to give
a nonzero v.e.v. of the metric $\tilde g$.
We have presented here a possible quantum mechanical origin for such a
potential.

Notice that the contribution of massless scalar fields to
$V_{\rm fin}$ is a $\theta$-independent constant.
One can check that this holds also in the case of
higher spin bosons, like the photon or the graviton.
Finally, the contribution of spinor fields to $V_{\rm fin}$,
is also expected to be of the form (4.9).
Therefore, even in presence of gauge fields and matter,
the minimum of $V_{\rm fin}$ will occur for $\theta$ multiple of the
identity.

The next step is the study of the renormalization group flow of this
minimum, as the characteristic IR scale of the theory changes;
this is of great relevance both in the quantization of gravity and in
cosmological problems, as we pointed out in [3, 9].
We plan to discuss this point in a separate publication.

\vfill\eject

\null
\centerline{\bf References}
\bigskip

\item{1.} R. Floreanini and R. Percacci, ``A multiplicative background
field method'', in {\it Gravitation theory and Geometric Methods in
Field Theory}, volume in honour of D. Ivanenko's 90th jubilee,
ed. V. Koloskov, Moscow (1994)
\smallskip
\item{2.} R. Floreanini and R. Percacci, Nucl. Phys. {\bf B 436},
141 (1994)
\smallskip
\item{3.} R. Floreanini and R. Percacci, The renormalization group
flow of the dilaton potential, SISSA 204/94/EP, {\tt hep-th 942181},
Phys. Rev. D, to appear
\smallskip
\item{4.} R. Percacci, in {\it Differential geometric methods in
Physics},
Proceedings of the XIIIth International Conference, held in Shumen,
Bulgaria, ed. H.D. Doebner and T.D. Palev, (World Scientific,
Singapore, 1986);\hfil\break
R. Percacci, {\it Geometry of nonlinear field theories},
(World Scientific, Singapore, 1986);
R. Floreanini and R. Percacci, Class. and Quantum Grav. {\bf 7},
975 (1990); {\it ibid.} {\bf 7}, 1805 (1990);
{\it ibid.} {\bf 8}, 273 (1991)
\smallskip
\item{5.} A. Komar, Phys. Rev. D {\bf 27}, 2277 (1983);
{\it ibid.} {\bf 30}, 305 (1984); J. Math. Phys. {\bf 26}, 831 (1985)
\smallskip
\item{6.} C.J. Isham, A. Salam and J. Strathdee,
Phys. Rev. D {\bf 3}, 1805 (1971); {\it ibid.} {\bf 5}, 2548 (1972)
\smallskip
\item{7.} R. Percacci, Nucl. Phys. {\bf B 353} 271 (1991)
\smallskip
\item{8.} R. Floreanini and R. Percacci, Phys. Rev. D {\bf 46}, 1566
(1992)
\smallskip
\item{9.}  R. Floreanini and R. Percacci, The heat-kernel and the
average effective potential, Phys. Lett. B, to appear

\bye